\documentclass[twocolumn]{aastex6}

\usepackage{graphicx}
\usepackage{verbatim}

\newcommand\teff{$T_{\mathrm{eff}}$}
\newcommand\logg{$\log\,g$}
\newcommand\loggf{$\log g_{F}$}
\newcommand\loggfeq{\log g_{F}}

\newcommand\mbol{$m_{\mathrm{bol}}$}

\begin{document}

\title{Quantitative Spectroscopy of Supergiants in the Local Group Dwarf Galaxy IC 1613: Metallicity and Distance}
\author{Travis A.\ Berger\altaffilmark{1}, Rolf-Peter Kudritzki\altaffilmark{1,4}, Miguel A.\ Urbaneja\altaffilmark{2}, Fabio Bresolin\altaffilmark{1}, Wolfgang Gieren\altaffilmark{3,5}, Grzegorz Pietrzy\'nski\altaffilmark{3,6}, \& Norbert Przybilla\altaffilmark{2}}
\altaffiltext{1}{Institute for Astronomy, University of Hawaii, 2680 Woodlawn Drive, Honolulu, HI 96822}
\altaffiltext{2}{Institut f\"ur Astro- und Teilchenphysik, Universit\"at Innsbruck, Technikerstr. 25/8, 6020 Innsbruck, Austria}
\altaffiltext{3}{Departamento de Astronom\'{\i}a, Universidad de Concepci\'on, Casilla 160-C, Concepci\'on, Chile} 
\altaffiltext{4}{University Observatory Munich, Scheinerstr. 1, D-81679 Munich, Germany}
\altaffiltext{5}{Millennium Institute of Astrophysics, Santiago, Chile}
\altaffiltext{6}{Nicolaus Copernicus Astronomical Center, Polish Academy of Sciences, ul. Bartycka 18, PL-00-716 Warszawa, Poland}

\begin{abstract}
We present the spectral analysis of 21 blue supergiant stars of spectral type late B to early A within the Local Group dwarf galaxy IC 1613 based on VLT-FORS2 low resolution spectra. Combining our results with studies of early B type blue supergiants we report a wide bi-modal distribution of metallicities with two peaks around [Z] $\sim$ --0.50 dex and [Z] $\sim$ --0.85 dex. The bi-modal distribution correlates with spatial location, when compared with column densities of neutral hydrogen in IC 1613. While the low [Z] objects appear in regions of relatively high ISM HI column densities or close to them, the high [Z] supergiants are found in the central HI hole almost devoid of hydrogen. This suggests varied chemical evolution histories of the young stellar populations in IC 1613. Utilizing the Flux-Weighted Gravity-Luminosity Relation (FGLR), we determine IC 1613's distance modulus as $m - M$ = 24.39 $\pm$ 0.11 mag. This value is in agreement within previous distance measurements using the near-infrared period-luminosity relationship of Cepheids and the tip of the red giant branch. 
\end{abstract}


\section{Introduction}
Because of their extreme brightness, individual blue supergiant stars can be studied spectroscopically in galaxies out to $\sim$10 Mpc. With multi-object spectrographs attached to 8 to 10m telescopes we can determine stellar parameters (\teff, \logg, metallicity, etc.) of individual objects, and, with further analysis, properties of the galaxies themselves, such as metallicities and metallicity gradients. This analysis is especially important when trying to understand nearby galaxies' chemical evolution and star formation history \citep{Kudritzki2015, Kudritzki2016, Bresolin2016}. In addition, high precision distances to these galaxies can be obtained through the flux-weighted gravity-luminosity relation (FGLR) method \citep{Kudritzki2003}, which uses the blue supergiant temperatures and gravities obtained via spectral analysis \citep{Kudritzki2008, Kudritzki2016}.

IC 1613 is a dwarf irregular galaxy and a member of the Local Group at a distance of about 720 kpc \citep{Pietrzynski2006}. Its metallicity and chemical evolution have been the subject of controversy. Photometry of the intermediate age population resulted in a logarithmic metallicity relative to the sun of [Z] $\sim$ --1.3 dex \citep{Cole1999, Rizzi2007}. \cite{Kirby2013}, through the analysis of red giant spectra, report  [Fe/H] = --1.19 $\pm$ 0.01 dex. From the study of H II regions, authors report oxygen abundance values relative to the sun ([O/H]) ranging between --1.1 and --1.0 dex \citep{Kingsburgh1995, Lee2003, Bresolin2007}. Nonetheless, quantitative optical spectroscopy of blue supergiants of early spectral types B0 and B5 (BSG from here on) led to higher oxygen abundances of [O/H] = --0.79 $\pm$ 0.08 dex \citep{Bresolin2007} and to a metallicity (based on the $\alpha$-elements silicon, magnesium and oxygen) of [Z] = --0.82 $\pm$ 0.04. Most recent work by Camacho (2017, PhD thesis, Universidad de La Laguna, Spain) confirms this result (see also Garcia et al., 2018, to be submitted). Unfortunately, the optical analysis of these BSG spectral types does not allow us to determine reliable metallicities based on iron group elements. However, \cite{Garcia2014} and \cite{Bouret2015} investigated HST UV spectra of O-stars and concluded that the iron abundance of these objects may be even somewhat higher, [Z] $\sim$ --0.7 dex. Such results have led to speculation that the ratio of $\alpha$ to iron elements ([$\alpha$/Fe]) of the young stellar population in this galaxy is smaller than the solar value, as already indicated in \cite{Tautvaisiene2007}. In this paper, the authors investigated three red supergiants of spectral type M in IC 1613 and found [Z] $\sim$ --0.67 dex and [$\alpha$/Fe] $\sim$ --0.1. If the higher metallicity based on the inclusion of iron group elements were confirmed, this would have interesting repercussions for the evolution history of this galaxy. It would also solve the puzzle that the observed massive stellar winds in this galaxy seemed to be too strong for the very low metallicity as derived from H II regions \citep{Tramper2011, Tramper2014}.

In this paper, we analyze ESO VLT FORS2 high quality ($S/N$ $\sim$ 70 or better) spectra of 21 late B--early A supergiants (ASG from here on). The major motivation for this work is to obtain more information about the metallicity of the young stellar population in IC 1613. In particular for the cooler objects among the ASG, optical metal line spectra are dominated by iron group elements, which will yield metallicities also based on iron group elements. We also determine the distance to IC 1613 using the FGLR method and compare with Cepheid and tip of the red giant branch (TRGB) distances to test the FGLR at low metallicity.

\section{Observations and Analysis Method}
\subsection{Spectra and Photometry}
The ASG observations took place on the nights of 2003 October 26 and 27 as part of the comprehensive spectroscopic survey of massive stars in IC 1613 by \cite{Bresolin2007}. The European Southern Observatory Very Large Telescope UT4 with the Focal Reducer and Low Dispersion Spectrograph 2 (FORS2) was used and three fields were observed with 19 movable slitlets each 1'' wide and 21'' long. The spectra cover the range from 3700--5900\AA\ with a spectral resolution of 5\AA\ (FWHM). For all details of the observations and their reductions we refer to \cite{Bresolin2007}, which also contains a discussion of the spectra including spectral classification.

In addition, as already discussed in the introduction, \cite{Bresolin2007} carried out a quantitative spectral analysis of nine early-type BSG but did not include the cooler objects in their analysis. In this paper, we now concentrate on the analysis of remaining cooler ASG. The list of objects studied is given in Table \ref{tab:targ}. The object identification is taken from \cite{Bresolin2007}. Three objects, which were originally included in our spectral analysis, are not listed in Table \ref{tab:targ}: objects C4 and C13 have stellar gravities (\logg) too high for the grid of non-LTE spectra used for the analysis (see below) and object A10 has an effective temperature (\teff) too large for the model grid. 

Our analysis also requires photometry in order to determine interstellar reddening, extinction and apparent bolometric magnitudes, which are then used for a determination of distance through the FGLR method. For this purpose we use the B- and V-band photometry obtained by \cite{Garcia2009} with the Wield Field Camera at the La Palma Isaac Newton Telescope. The photometric data are also provided in Table \ref{tab:targ}.

\begin{figure*}[htp!]
\centering
\includegraphics[scale=0.57]{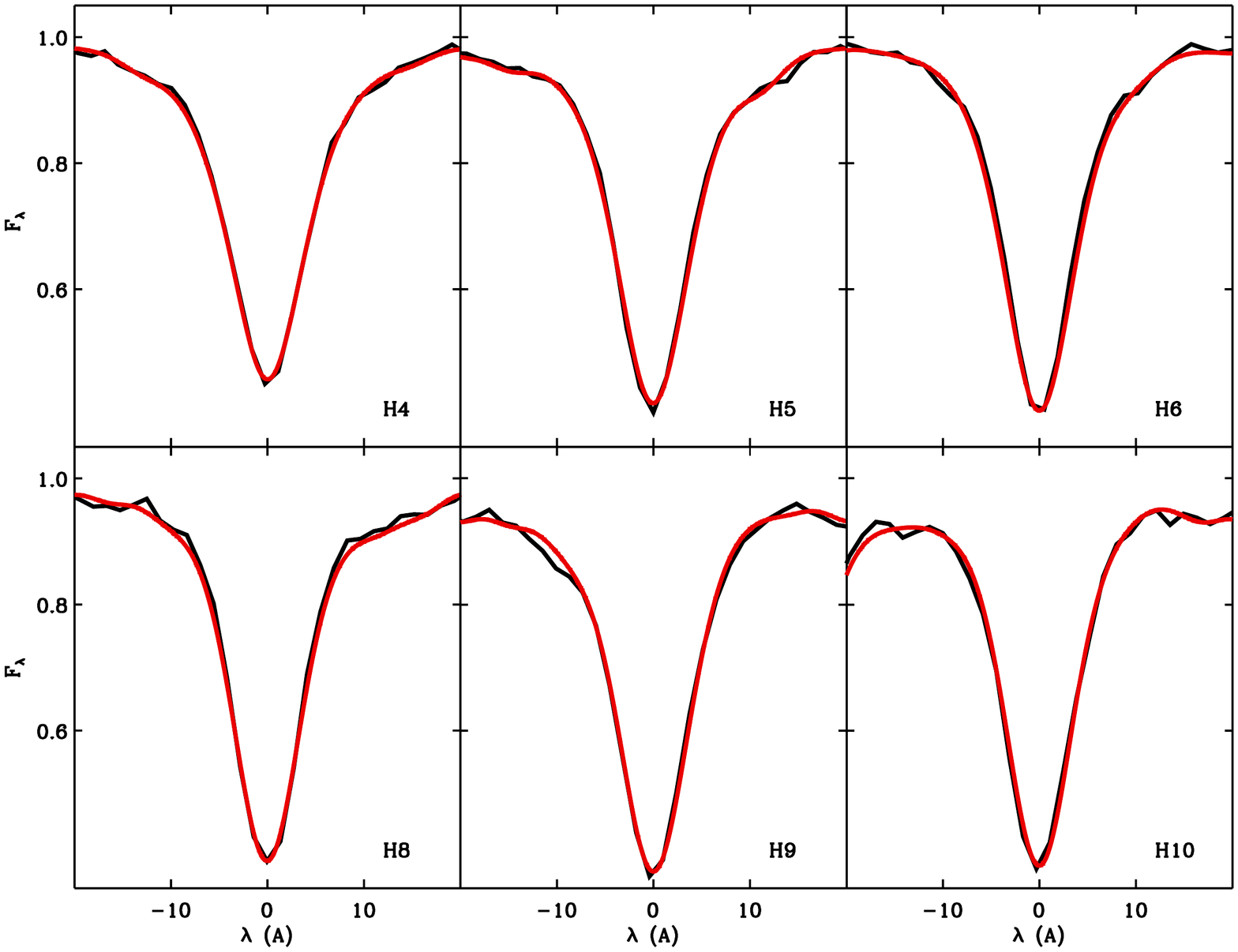}
\includegraphics[scale=0.57]{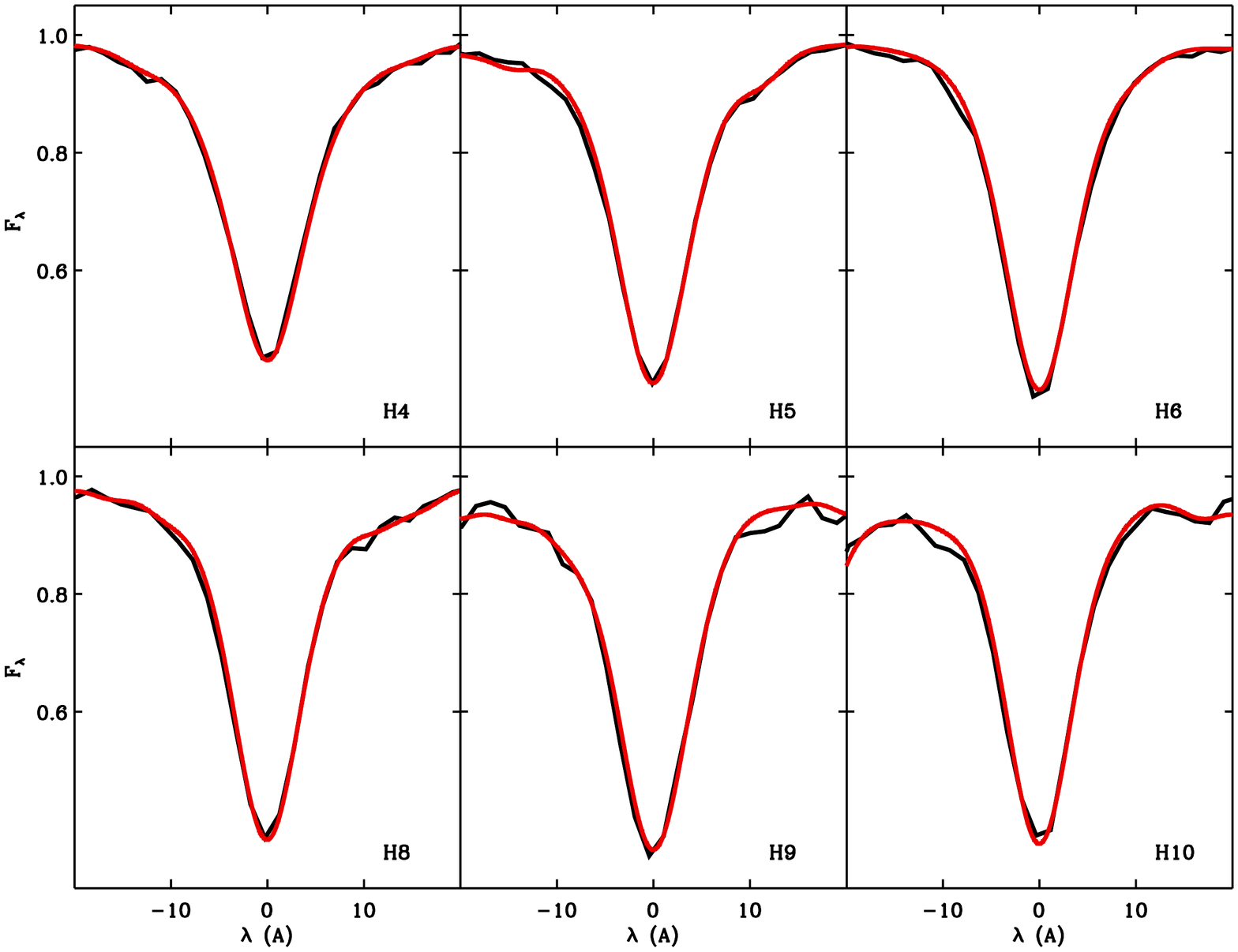}
\caption{Balmer line fits for targets A1 (top) and C1 (bottom). The red curve represents the normalized object spectrum, while the black curve is a normalized synthetic spectrum. The synthetic spectra were computed for \teff\ = 8750 K, \logg\ = 1.90 dex, [Z] = --0.50 dex (target A1) and \teff\ = 8300 K, \logg\ = 1.60 dex, and [Z] = --0.85 dex (target C1). The x-axis is the displacement from the line center measured in \AA.} \label{fig:balmComp}
\end{figure*}

\begin{figure*}[ht!]
\centering
\includegraphics[scale=0.5]{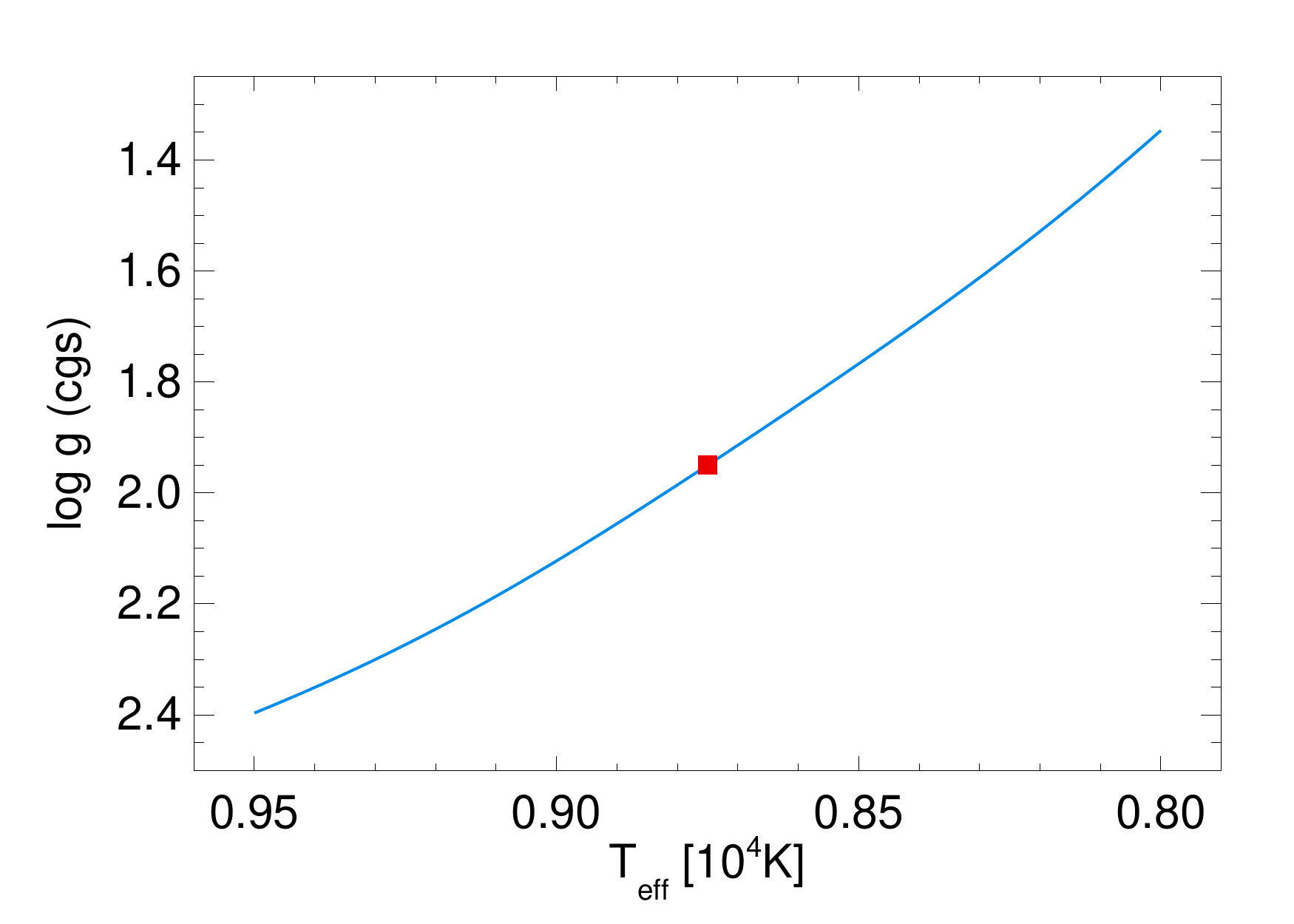}
\includegraphics[scale=0.5]{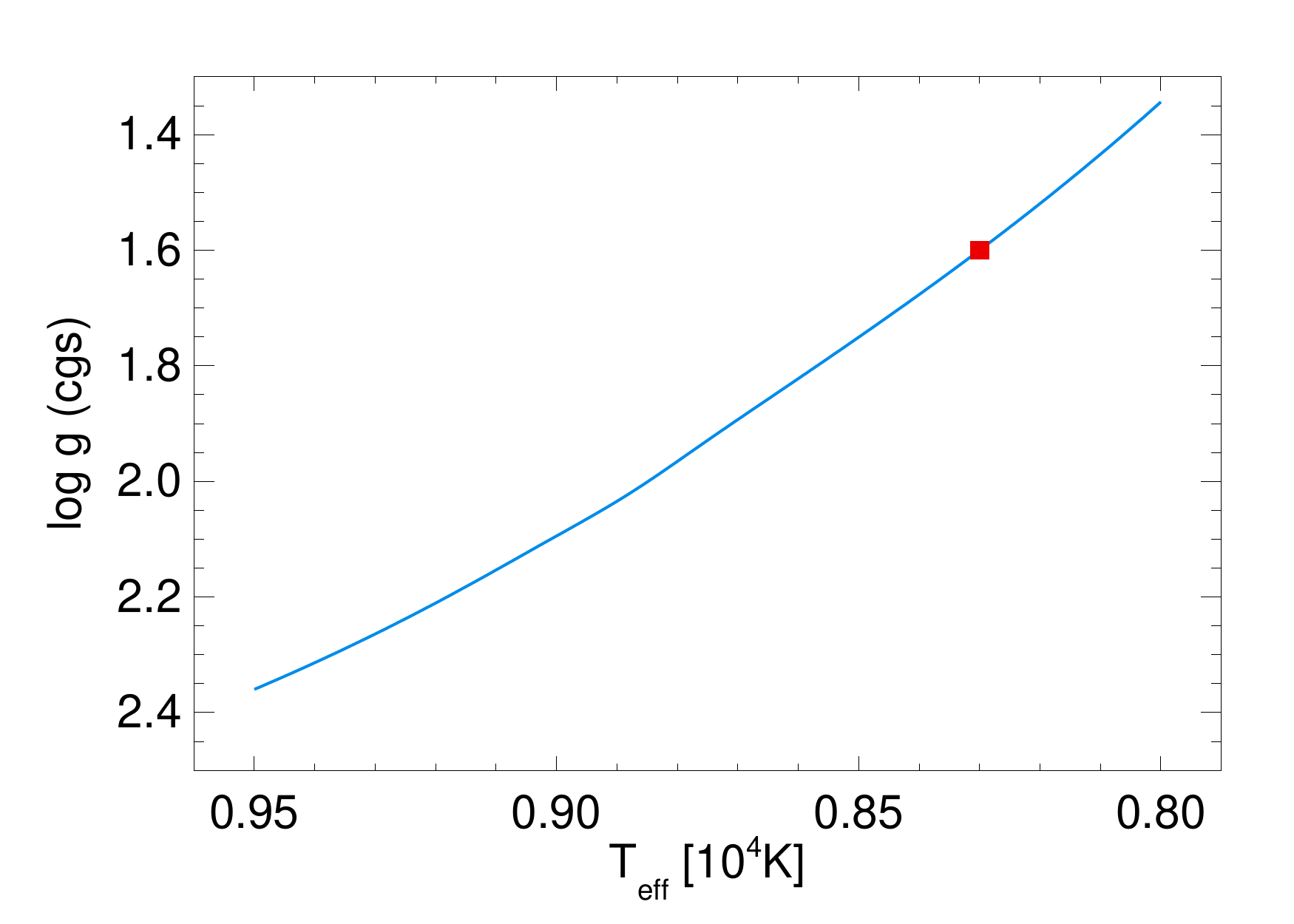}
\caption{Fit curves of the Balmer lines in the gravity-temperature plane for targets A1 (left) and C1 (right). Along these curves the calculated Balmer lines agree with the observations. The red squares correspond to the fit parameters of Figure \ref{fig:balmComp}.} \label{fig:balmiso}
\end{figure*}

\begin{figure*}[ht!]  
\centering
\includegraphics[scale=0.59]{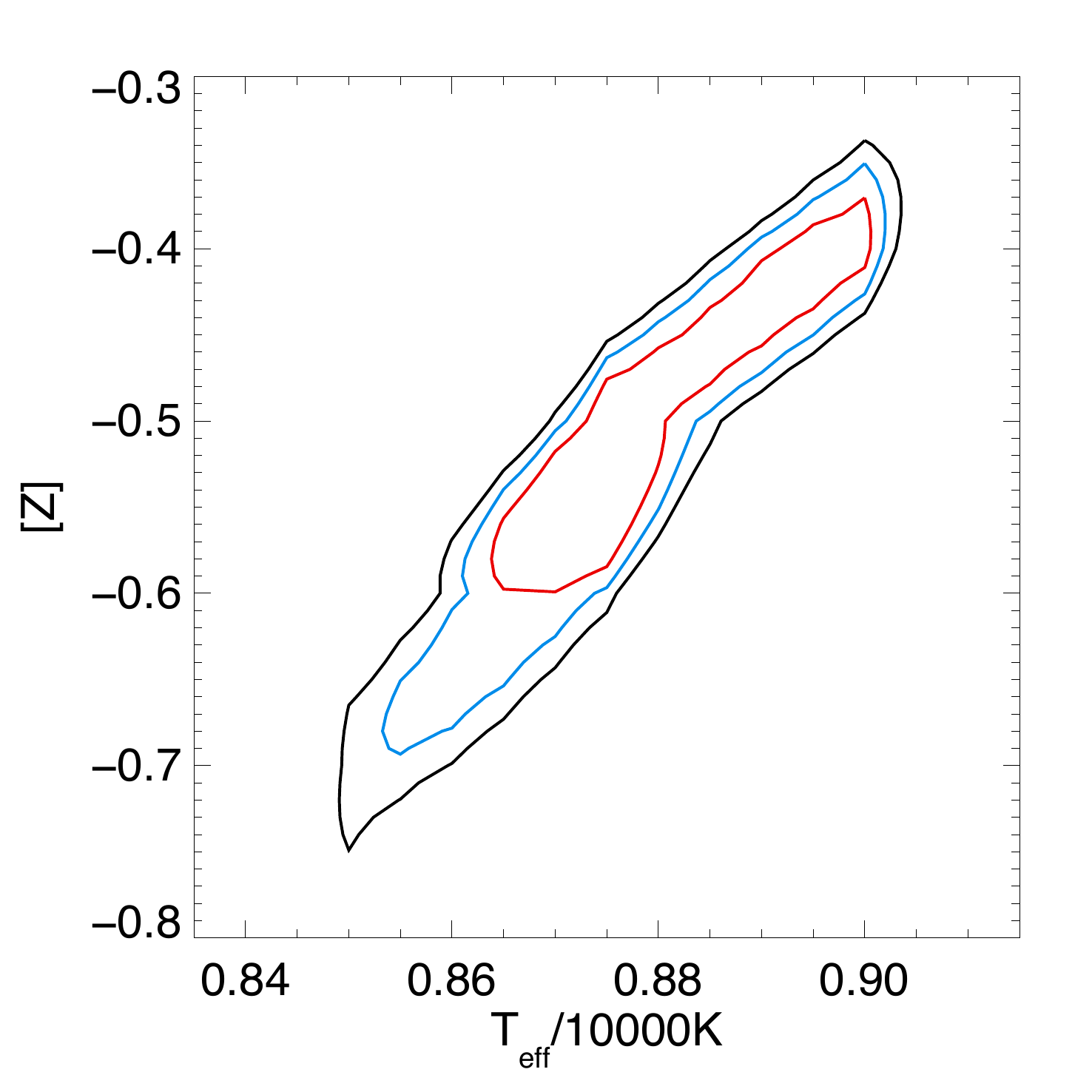}
\includegraphics[scale=0.59]{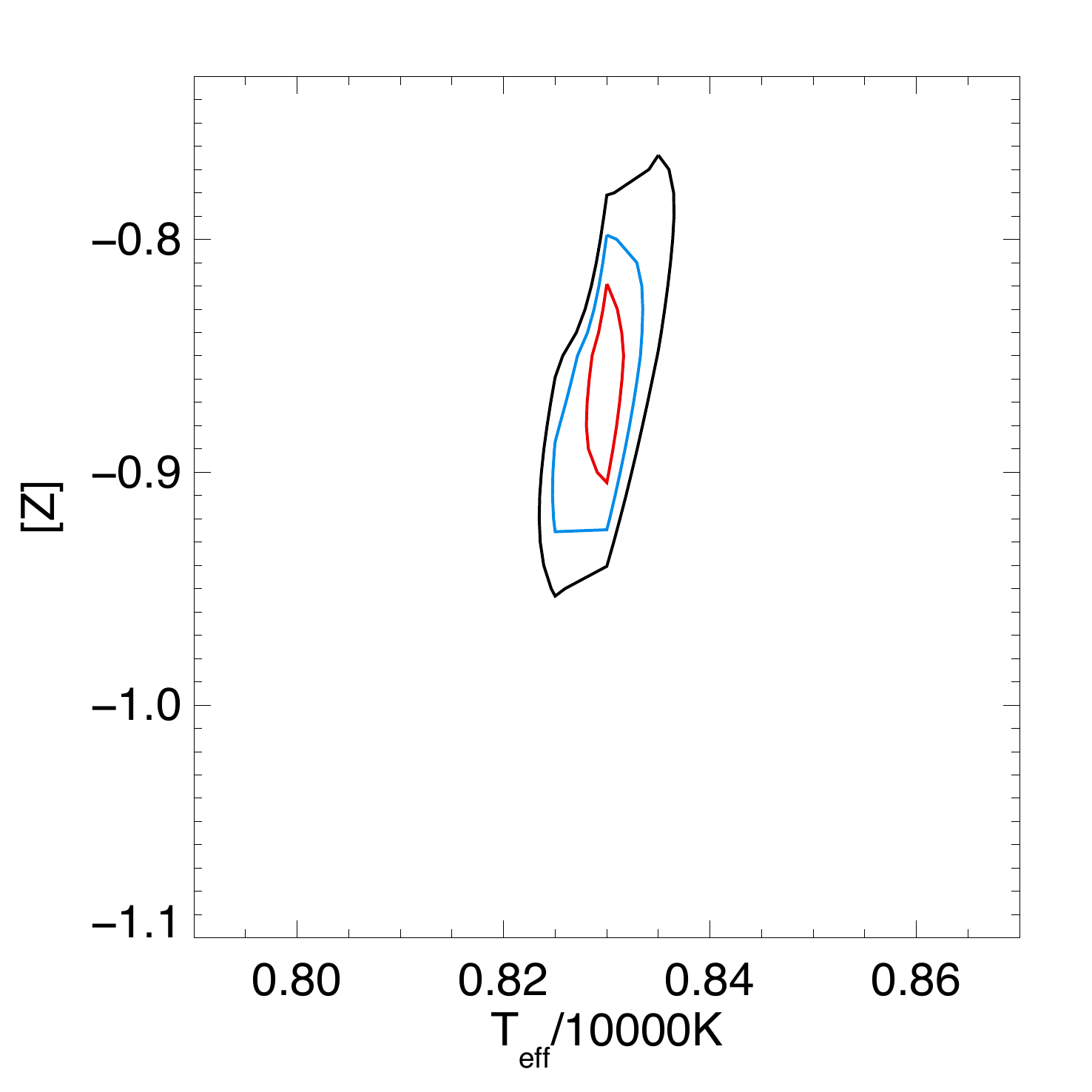}
\caption{Determination of \teff\ and [Z] from isocontours $\Delta \chi^2$ in the metallicity-temperature plane obtained from the comparison of synthetic with observed spectra (see text). Plotted are $\Delta \chi^2$ = 3 (red), 6 (blue), and 9 (black), respectively. $Left$: target A1; $right$: Target C1.} \label{fig:chisq}
\end{figure*}

\begin{figure*}[htp!]
\centering
\includegraphics[scale=0.62]{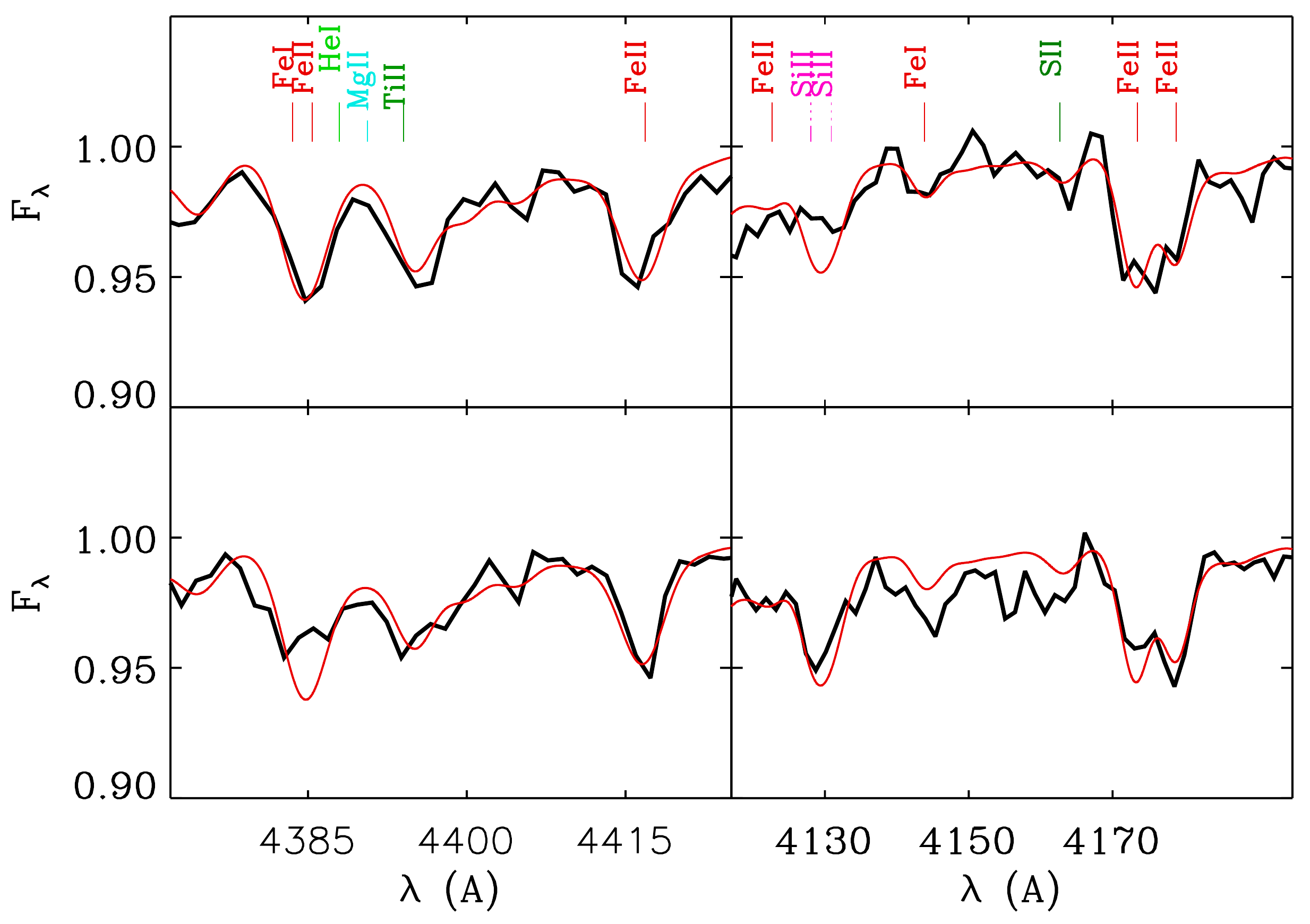}
\includegraphics[scale=0.62]{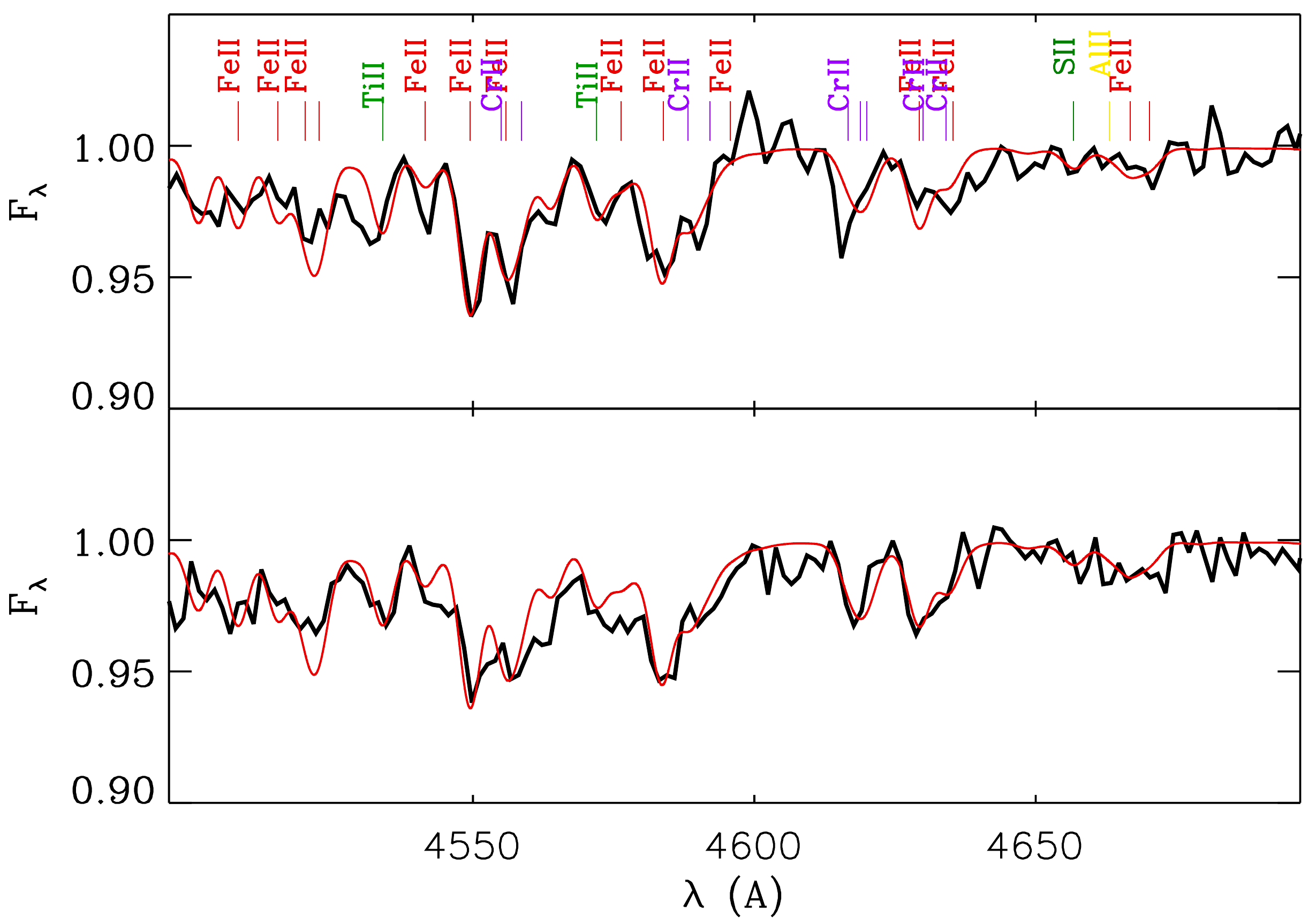}
\caption{Examples of metal line fits for targets C1 (top) and A1 (bottom) in three spectral windows using the final model parameters. Each of the metal lines are labeled accordingly. The black curve represents the normalized object spectrum, while the red curve is a normalized synthetic spectrum.} \label{fig:metfit}
\end{figure*}

\subsection{Analysis Method}
We follow the same procedure described in \citet{Kudritzki2013,Kudritzki2016} and \cite{Hosek2014} to determine the stellar parameters \teff\ and \logg\ and stellar metallicity ([Z]), defined as [Z] = log Z/Z$_{\odot}$, where Z$_{\odot}$ is the solar metallicity. We utilize a grid of LTE line-blanketed atmospheres with synthetic spectra obtained with non-LTE line formation calculations using the elaborate atomic models described in \cite{Przybilla2006}. The grid is comprised of temperatures from 7900--15000 K, while \logg\ varies between 0.8 and 3.0 dex in cgs units (the exact upper and lower limits depend on \teff). [Z] ranges from --1.30 to 0.50 dex. Details of the grid are described in \citet{Kudritzki2008,Kudritzki2012}. Figure 1 in \cite{Kudritzki2008} illustrates the extent and spacing of the model atmosphere grid.

The synthetic spectra are then compared with the normalized observed spectra to obtain stellar parameters and metallicity. In a first step we use the observed Balmer lines H$_{4,5,6,7,8,9,10}$ to constrain \logg\ as a function of \teff. At fixed values of \teff\ we determine the gravity which fits the Balmer lines best. Figure \ref{fig:balmComp} provides some examples of the best-fit model spectra for different targets' Balmer lines. \logg\ is usually determined with an accuracy of 0.05 to 0.10 dex at a fixed value of \teff\ (Table \ref{tab:targ} provides the corresponding values for the individual targets). Since the strength of the model Balmer lines decreases with increasing temperature, the \logg\ fit values increase with temperature. Figure \ref{fig:balmiso} shows two examples of the relationships between \teff\ and \logg\ obtained in this way.  Next, we move along the gravity-temperature relationship determined from the Balmer lines and compare observed and calculated flux as a function of metallicity in 11 spectral windows, which are dominated by metal lines. We assess the quality of the fit by calculating a $\chi^2$ value at each \teff\ and [Z]. Then, we calculate the minimum of $\chi^2$ and isocontours in $\Delta \chi^2$ around the minimum, which then provide a determination of \teff\ and [Z] together with an estimate of the uncertainties. We obtain \logg\ from the gravity-temperature relationship determined in the first step. Figure \ref{fig:chisq} shows the corresponding $\Delta \chi^2$ isocontours for two targets. We emphasize that through the $\chi^2$ fit of many spectral lines in many spectral windows, the resultant stellar metallicity reflects the contribution from many elements, including the iron peak and $\alpha$-elements. Figure \ref{fig:metfit} shows the fits obtained for some of metal line windows for two targets.

Once temperature, gravity and metallicity are obtained, we use the synthetic non-LTE spectral energy distribution (SED) of the final model to calculate intrinsic colors B-V and determine the interstellar reddening E(B-V). By assuming R$_{V}$ = A$_{V}$/E(B-V) = 3.3 for the ratio of extinction to reddening we then constrain the value of extinction A$_{V}$. Apparent bolometric magnitudes m$_{bol}$ for each star are then obtained from the de-reddened V-magnitudes and the bolometric correction BC calculated from the SED. The values of \teff, \logg, [Z], E(B-V), BC and m$_{bol}$ for each object analyzed are given in Table \ref{tab:targ}. The table also contains the stellar flux weighted gravity defined as \loggf\ = \logg\ -- 4log\teff\ + 16, which we will use for a determination of distances. The results compiled in the table will be discussed in the following sections.

\section{Blue supergiants of early spectral type}
For the discussion of metallicities and the determination of distance we will include the BSG in IC 1613 studied by \cite{Bresolin2007}, which were observed with the same spectroscopic setup at the same time as our ASG and then analyzed in detail using non-LTE model atmospheres. Most recently, Camacho (2017, PhD thesis) and Garcia, Camacho, Herrero et al. (2018, to be published, see Section 1) have extended this work using VIMOS multi-object spectroscopy obtained with the ESO VLT. They re-analyzed most of the targets investigated by \cite{Bresolin2007} but also added additional objects to the sample. For the objects in common the results are very similar to \cite{Bresolin2007}. In our comparison we use the results by Camacho, but add two more targets (A7 and A18) from \cite{Bresolin2007}, which were not included in the Camacho study. The work by Camacho uses the same photometry as we have been using for the determination of reddening and bolometric magnitudes \citep{Garcia2009}. For the two targets added from \cite{Bresolin2007} we have re-determined reddening and bolometric magnitudes again using the photometry by \cite{Garcia2009} for consistency.

\begin{figure}[t]
\includegraphics[scale=0.39]{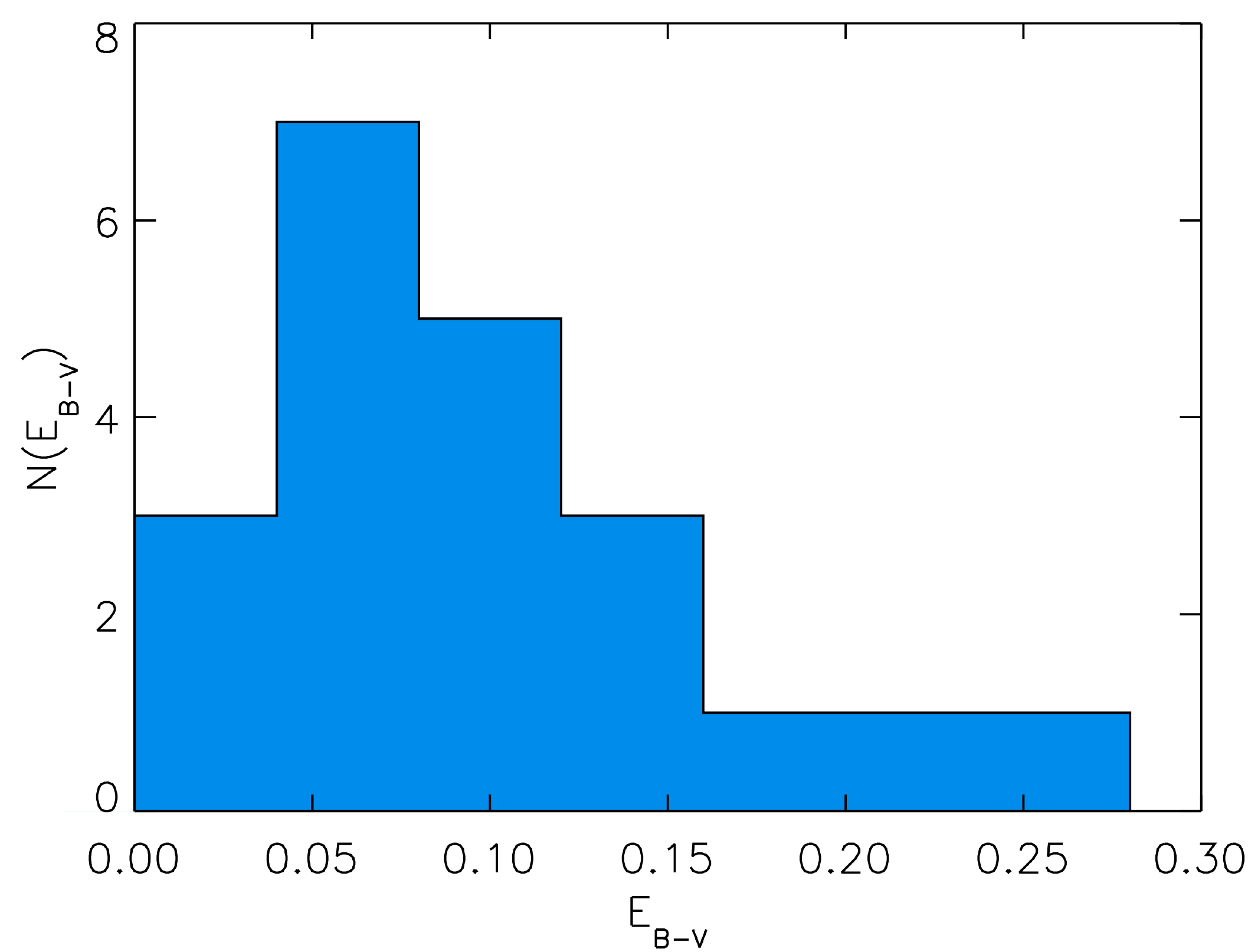}
\caption{Histogram of the distribution of interstellar reddening E(B-V) of the ASGs in Table \ref{tab:targ}.} \label{fig:nebv}
\end{figure}

\section{Interstellar reddening}
The measurement of interstellar reddening is crucial for the determination of accurate distances. As is well known, purely photometric stellar distance determination methods can be affected significantly by interstellar dust. Therefore, a fundamental advantage of the quantitative spectral analysis of blue supergiant stars is that it provides direct information about interstellar reddening. For an actively star forming galaxy such as IC 1613, the massive stars can be imbedded in a dusty environment so we expect a wide range of interstellar reddening. Figure \ref{fig:nebv} displays the distribution of E(B-V) among the ASG targets of Table \ref{tab:targ}. Indeed we find a wide distribution with reddening values larger than the Milky Way foreground value of 0.03 mag \citep{Pietrzynski2006}. The average value is 0.10 mag. A similar average value, 0.09 mag, has been found by \cite{Pietrzynski2006} by comparing distance moduli obtained from period--luminosity relationships of Cepheids at different wavelengths ranging from the NIR to optical wavelengths, but no individual values could be determined. The distribution of reddening values agrees well with the comprehensive photometric study by \cite{Garcia2009}.

\section{Evolutionary status}
\begin{figure}[t]
\includegraphics[scale=0.60]{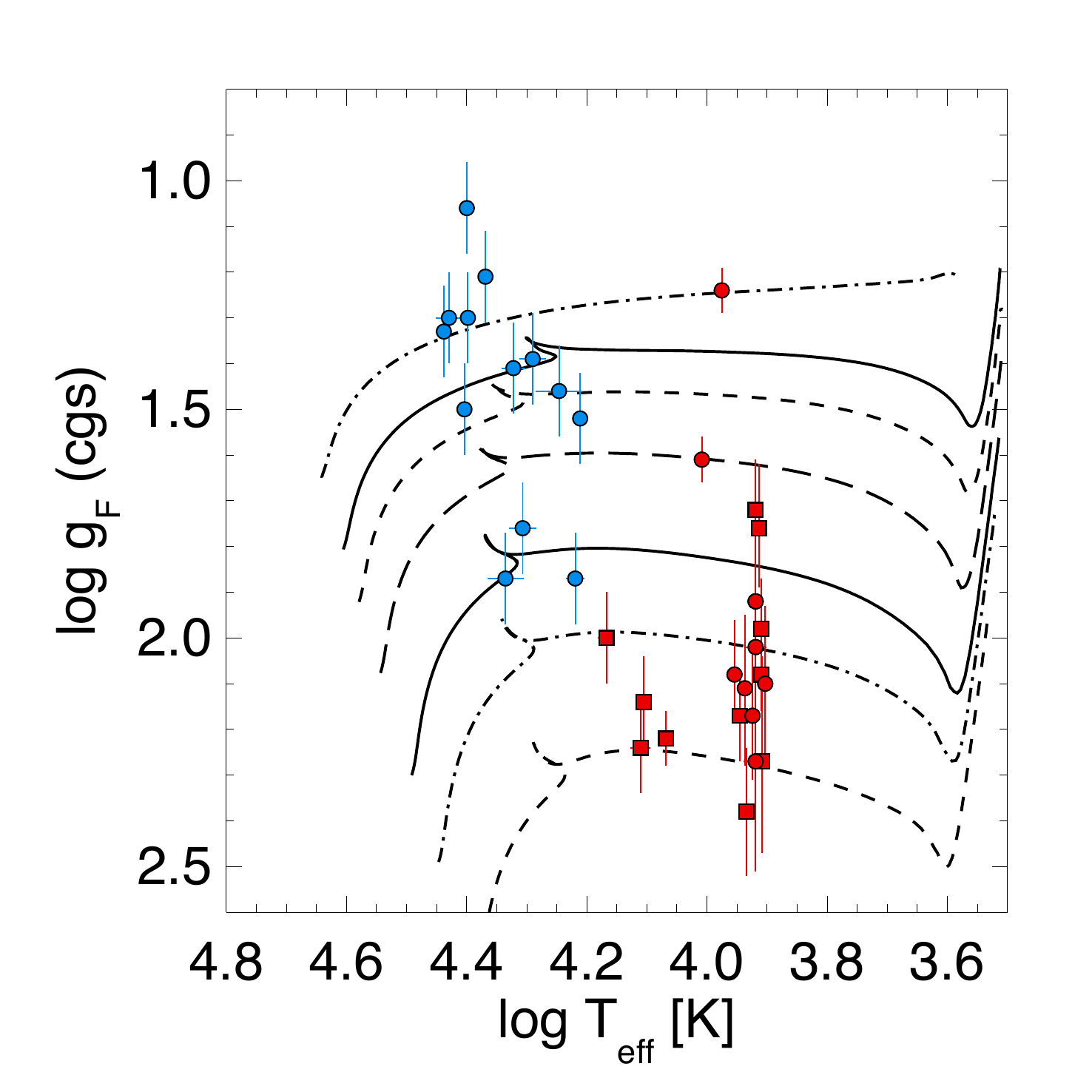}
\caption{Spectroscopic Hertzspung--Russell diagram of blue supergiant stars in IC 1613. Plotted is the flux weighted gravity \loggf\ versus the logarithm of \teff. The ASG from this study are shown as red squares for metallicities [Z] larger than --0.7 and as red circles for metallicities lower than or equal to --0.7 (see discussion in \S6). The BSG of early type B are plotted as blue circles. Their metallicities are all smaller than --0.7. Evolutionary tracks at SMC metallicity by \cite{Brott2011} calculated for initial main-sequence masses of 9, 12, 15, 20, 25, 30, 40 M$_{\odot}$ (from the bottom of the figure to top) are also displayed. The calculations of the tracks include the effects of rotation with initial main-sequence rotational velocities of about 180 km/sec.} \label{fig:shrd}
\end{figure}

The spectroscopic determination of the ASG stellar parameters summarized in Table \ref{tab:targ} and of the BSG studied by Camacho (2017) and \cite{Bresolin2007} allow us to discuss the evolutionary status of these blue supergiants. A good direct approach for such a discussion is the ``spectroscopic Hertzsprung-Russell diagram'' (sHRD) introduced by \cite{Langer2014}. The sHRD displays flux-weighted gravities against effective temperature and is equivalent to the classical Hertzsprung-Russell diagram (HRD) except that it is independent of the distances adopted. Figure \ref{fig:shrd} shows the sHRD for our objects and demonstrates very nicely that the blue supergiants are in an evolved stage of stellar evolution. The hotter BSG objects are somewhat younger and closer to the original zero-age-main-sequence and the cooler ASG objects are more advanced in their evolution. We also see that most of the cooler ASG objects have lower initial zero-age-main-sequence masses than the hotter BSG counterparts in the sHRD. This is the result of a selection effect related to the V-band magnitude limit of the spectroscopic observations and the much higher bolometric correction of the earlier spectral types. More broadly, Figure \ref{fig:shrd} demonstrates that the early type and late type objects belong to the same stellar population, although there is a small age difference of $\sim$20 million years between the hotter, more massive objects and the cooler, less massive objects according to the stellar evolution calculations by \cite{Brott2011}, which are also illustrated in the figure.

\begin{figure}[t]
\centering
\includegraphics[scale=0.75]{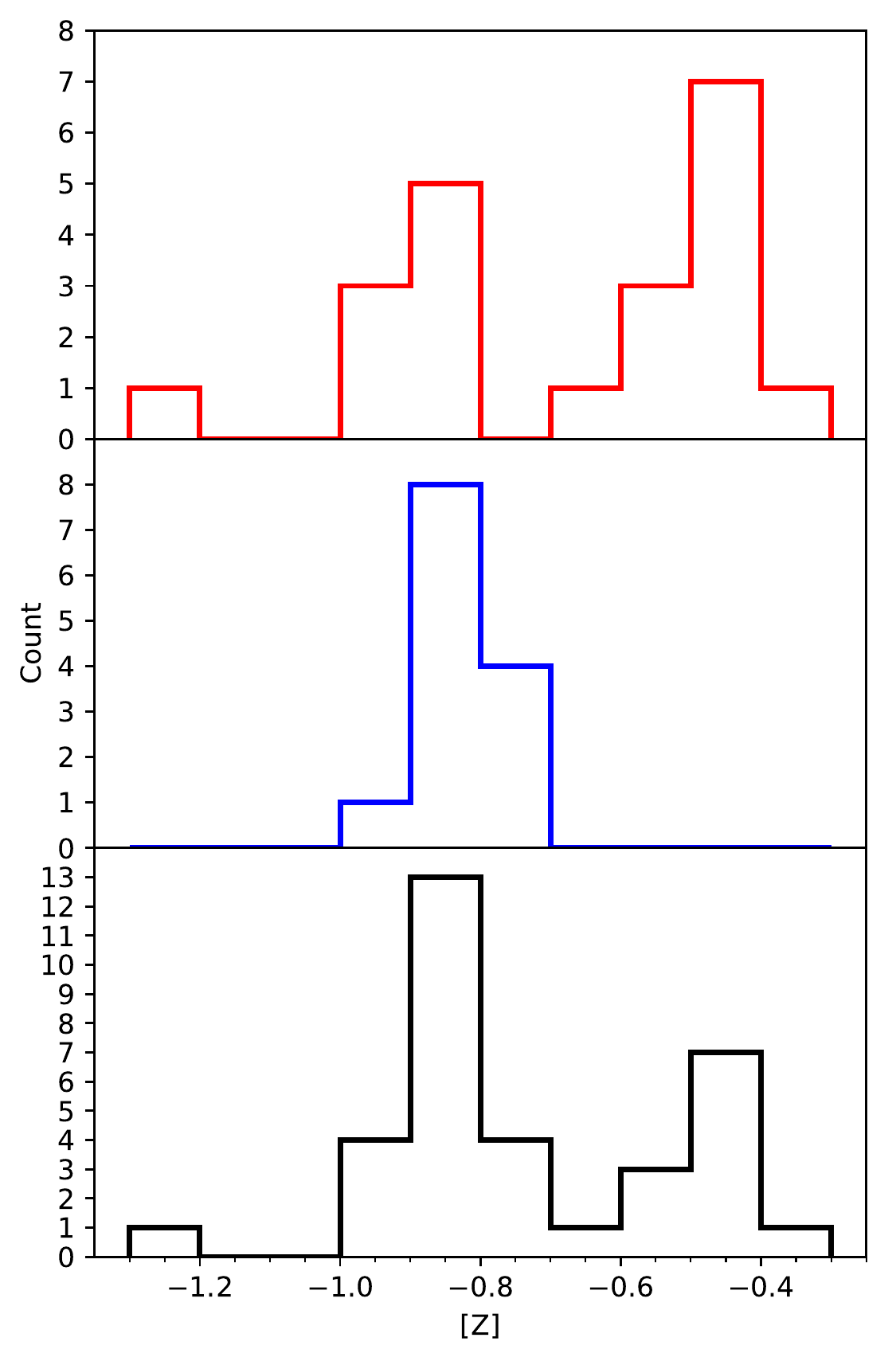}
\caption{Distribution of blue supergiant metallicities [Z] in IC 1613. $Upper$: the 21 ASG, $middle$: the 13 early type BSG, and $bottom$: the total sample.} \label{fig:zhisto}
\end{figure}

\begin{figure*}[t]
\centering
\includegraphics[scale=0.63]{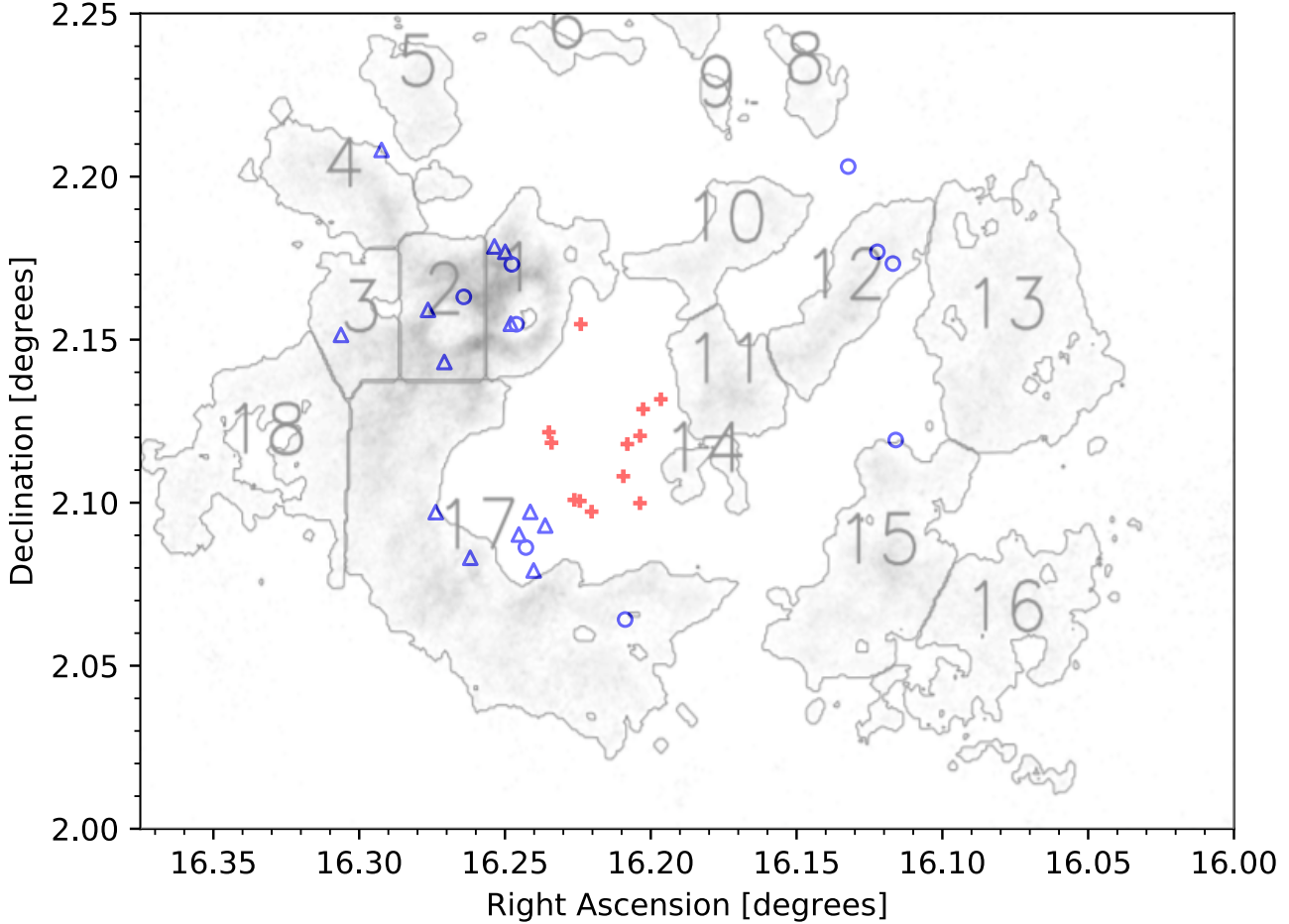}
\caption{Spatial distribution of blue supergiants in IC 1613. The 21 ASG are plotted either as blue circles ([Z] $\leq$ --0.70) or red crosses ([Z] $\geq$ --0.70). The 13 BSG, which all have metallicities lower than [Z] = --0.7, are displayed as blue triangles. In addition, we overplot the HI regions from \cite{Silich2006} in translucent grey. Darker color corresponds to higher HI column densities.} \label{fig:metdist}
\end{figure*}

\section{Metallicity}
The average metallicity of the total sample of all blue supergiants, ASG and BSG, is [Z] = --0.69 $\pm$ 0.24 dex. This value is comparable to the average metallicities found for three O-stars and three M-supergiants by \cite{Bouret2015} and \cite{Tautvaisiene2007}, respectively. This result is consistent with stellar evolution theory, since O-stars and M-supergiants are the direct precursors and successors of blue supergiants. For comparison, the average metallicity obtained from the analysis of 125 red giant stars within IC 1613 is [Fe/H] = --1.19 $\pm$ 0.01 dex \citep{Kirby2013}. While the metallicity derived from the analysis of red giants is significantly smaller than the blue supergiant value, red giants are, on average, significantly older than blue supergiants. Therefore, we expect such discrepancies in metallicity between these two populations.

Figure \ref{fig:zhisto} shows the distribution of metallicities [Z]. The result is surprising, because the [Z] distributions of the ASG and the hot BSG are significantly different. While the hot objects are narrowly distributed with a clear peak at [Z] $\sim$ --0.85, the ASG show an additional component peaking a higher metallicity of [Z] $\sim$ --0.50. We stress that this difference can hardly be the result of different stellar age.

The bimodal ASG metallicity distribution is an interesting result in need of further investigation. For [Z] $\leq$ --0.7 it is similar to the hot BSG [Z] distribution, but there is an additional component with metallicities higher than --0.7. In Figure \ref{fig:metdist} we investigate whether the occurrence of this higher metallicity component is related to the spatial location within the galaxy. We plot both groups, BSG and ASG, but distinguish between metallicities higher and lower than [Z] = --0.7. The result is striking. The higher metallicity objects are concentrated in the central region of the galaxy and spatially separated from supergiants of lower metallicity. Most interestingly, there is also an indication of a correlation with the existence of neutral HI ISM gas. We overplot the extent of the HI regions observed by \cite{Silich2006} within IC 1613 as translucent, grey contours where darker colors indicate higher column densities. This reveals a significant pattern: the higher metallicity objects appear to cluster in the central areas of low HI column densities, while the lower metallicity objects typically appear within the ring-like regions of higher HI column densities or close to them.

From the viewpoint of chemical evolution the interpretation is straightforward. In the HI gas-depleted regions, the star formation process has proceeded more rapidly and has consumed the HI gas to form stars, which, through supernova explosions and stellar winds, have enriched the metallicity in these regions. Thus, there appear to be two populations of ASGs in IC 1613:  those born from the higher metallicity ashes of stars in the center of the galaxy, and those born from the low metallicity HI region gas and dust in the outer portions of the galaxy. Although both the metal-poor HI regions and the central, metal-rich portions of the galaxy show signs of recent star formation, they remain separated in their metallicities, indicative of different chemical evolution histories.

\section{FGLR and Distance Determination}

\begin{figure}[t]
\centering
\includegraphics[scale=0.6]{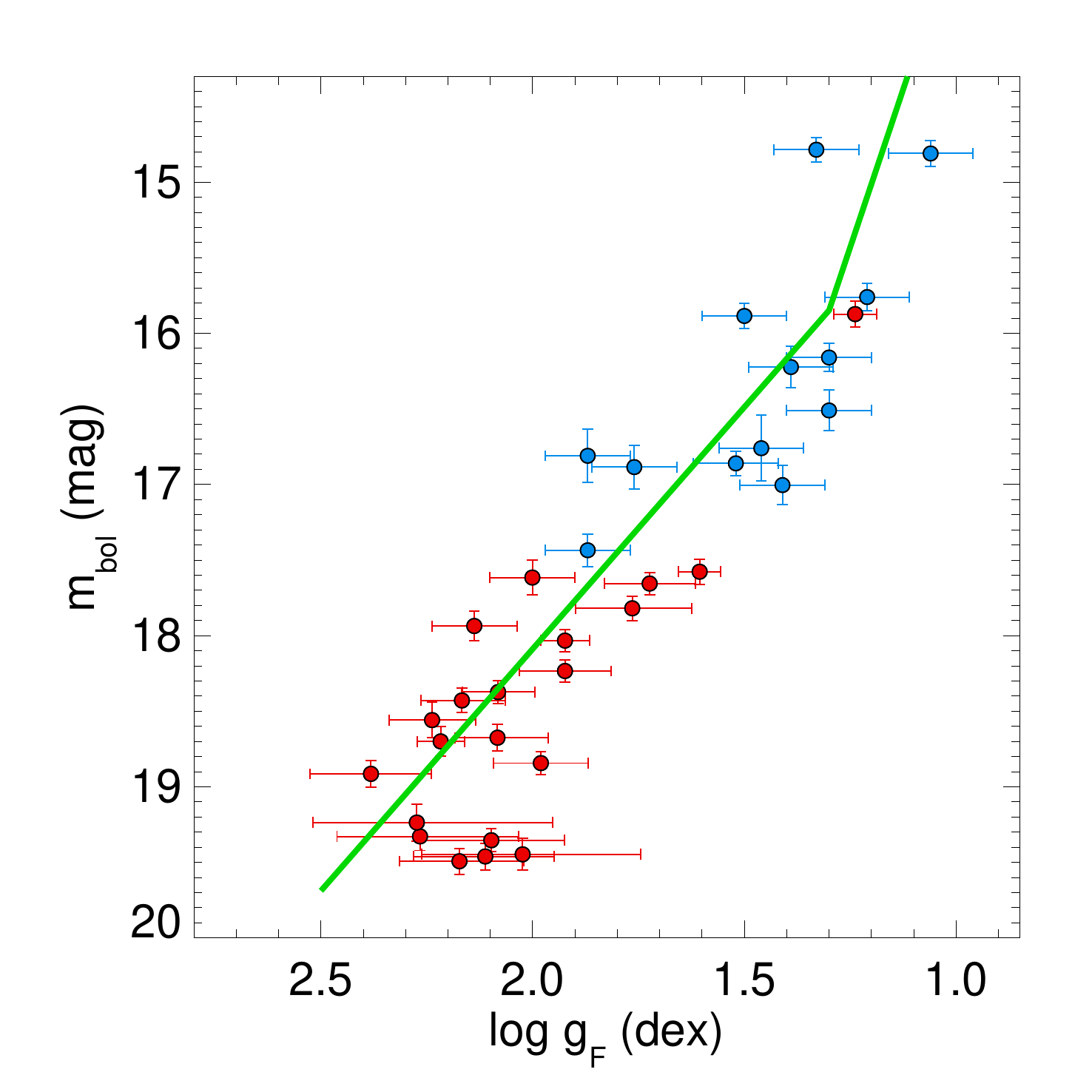}
\caption{FGLR for IC 1613. The green line represents the LMC calibration by \cite{Urbaneja2017} fitted to the data in a simple least square procedure to obtain the distance modulus. The red points correspond to ASGs, while the blue points represent the BSGs.} \label{fig:fglr}
\end{figure}

\citet{Kudritzki2003,Kudritzki2008} have shown that the flux-weighted gravity \loggf~= \logg~-- 4log\teff~+ 16 of blue supergiant stars is tightly correlated with their absolute bolometric magnitude. As explained in their work, this is because blue supergiants cross the HRD at roughly constant luminosity and mass as they evolve away from the main sequence (see the evolutionary tracks of Figure \ref{fig:shrd} for some examples). This correlation, the FGLR, qualifies blue supergiants as excellent distance indicators and has already been used to determine the distances to many galaxies \citep{Kudritzki2012,Kudritzki2014,Kudritzki2016,Bresolin2016,Hosek2014,Urbaneja2008,U2009}.

Combining the flux-weighted gravities and apparent bolometric magnitudes of the ASG and BSG, we obtain the observed blue supergiant FGLR of IC 1613 in Figure \ref{fig:fglr}. In order to determine the distance to IC 1613, we use the new FGLR calibration by \cite{Urbaneja2017} resulting from a detailed spectroscopic analysis of 90 blue supergiants in the Large Magellanic Cloud (LMC):

\begin{equation}
M_{\mathrm{bol}} = a(\loggfeq - 1.5) + b
\end{equation}
if $\loggfeq \geq \loggfeq^{\mathrm{break}}$, and
\begin{equation}
M_{\mathrm{bol}} = a_{\mathrm{low}}(\loggfeq - \loggfeq^{\mathrm{break}}) + b_{\mathrm{break}}
\end{equation}
if $\loggfeq \leq \loggfeq^{\mathrm{break}}$, with
\begin{equation}
b_{\mathrm{break}} = a(\loggfeq^{\mathrm{break}} - 1.5) + b,
\end{equation}
where $\loggfeq^{\mathrm{break}}$ = 1.30 dex, $a$ = 3.20 $\pm$ 0.08, $b$ = --7.90 $\pm$ 0.02 mag, and $a_{\mathrm{low}}$ = 8.34 $\pm$ 0.25.

With this LMC FGLR calibration, we calculate absolute magnitudes from the observed \loggf\ values. This yields individual distance moduli for each object. The distance is then determined from a weighted mean that accounts for the observational errors of \loggf\ and m$_{bol}$. We obtain a distance modulus of $m - M$ = 24.39 $\pm$ 0.11 (r) $\pm$ 0.05 (s) mag corresponding to a distance of 759 $\pm$ 43 kpc. Here, the systematic error accounts for the uncertainties of the calibration FGLR parameters. The green curve in Figure \ref{fig:fglr} represents the calibration FGLR \citep{Urbaneja2017} shifted to our determined distance.

The FGLR distance is in agreement with the value of $m - M$ = 24.29 $\pm$ 0.05 mag obtained by \cite{Pietrzynski2006} using NIR observations of Cepheids. \cite{Tammann2011} also used Cepheids to determine a distance relative to the Small Magellanic Cloud. Their distance modulus, $m - M$ = 24.38 $\pm$ 0.03 mag, was corrected for the new SMC distance found by \cite{Graczyk2014}. The FGLR distance also agrees with the Tip of the Red Giant Branch (TRGB) distance modulus, $m - M$ = 24.38 $\pm$ 0.10 mag, given by the Extragalactic Distance Database (EDD, http://edd.ifa.hawaii.edu, \citeauthor{Tully2009} \citeyear{Tully2009}) and with the distance modulus, $m - M$ = 24.37 $\pm$ 0.08 mag, obtained from the period luminosity relationship of carbon-rich Mira stars \citep{Menzies2015}.

We note that the scatter around the green FGLR fit curve in Figure \ref{fig:fglr} with $\sigma$ = 0.6 mag is larger than the typical values of $\sigma$ = 0.25--0.40 mag encountered for other galaxies. As the stellar evolution simulations of the FGLR by \cite{Meynet2015} indicate (compare their Figures 8 and 10), this might be an effect of the low metallicity at which the effects of rotational mixing are more pronounced and lead to a significant deviation from constant luminosity for rotating stars when they cross the HRD \citep{Georgy2013}. Conversely, the FGLR of blue supergiants in the metal poor galaxies WLM \citep{Urbaneja2008} and NGC~3109 \citep{Hosek2014} shows a scatter comparable to galaxies with higher metallicity, when a fit using the new calibration of Equations (1) to (3) is applied (Kudritzki et al., 2018, in preparation). We also note that the evolutionary tracks by \cite{Brott2011} for low metallicity stay at roughly constant luminosity when crossing the HRD and do not show the behavior found in the work by \cite{Georgy2013}.

\section{Discussion and Future Work}

The results obtained in the combination of this work and of the investigation by Camacho and collaborators indicate a spatially inhomogeneous chemical evolution and star formation history of IC 1613. This is in agreement with the detailed study by \cite{Bernard2007} based on resolved-star V- and I-band photometry carried out with the 2.5m Isaac Newton Telescope. This study revealed that over the last 0.15--2 Gyr the star formation rate in the central region, now void of ISM neutral hydrogen, has been significantly higher than in the surrounding ring-like structure of higher hydrogen density. Then, over the last 100 Myr, very likely because of the consumption of central ISM hydrogen gas, star formation in the surrounding ring has become larger than in the central region. The reduced ISM hydrogen density in the center may simply be the result of enhanced star formation. However, dynamical processes such as stellar winds and supernovae explosions may have contributed to the reduced ISM hydrogen density, too, as the complex velocity and density structure of IC~1613 indicates \citep{Valdez2001, Lozinskaya2003, Silich2006}.

Enhanced star formation, which consumes interstellar gas to form stars, accelerates chemical evolution through the death of massive and intermediate mass stars. The metallicity of the ISM and the young stellar population is then a function of the ratio of stellar mass to gas mass. The HI surface density in the central cavity is about a factor of two lower than the average surface density in the surrounding ring structure, as the isocontours shown in \cite{Lake1989} indicate. On the other hand, the stellar surface density is slightly enhanced in the central region \citep{Bernard2007} so that the ratio of stellar mass to gas mass is higher by about a factor of three to four in the center compared to the ring. As chemical evolution models show, see for instance \cite{Kudritzki2015} Figure 1, such a difference in the ratio of stellar to gas mass can easily explain the difference in metallicity of 0.35 dex (between the high and low metallicity ASG) as an effect of chemical evolution caused by enhanced star formation.

Of course, this interpretation needs to be checked independently, ideally through spectroscopy of other types of massive stars or HII-regions in the central neutral hydrogen cavity. So far, all the O-stars  \citep{Garcia2014, Tramper2014, Bouret2015} and M supergiants \citep{Tautvaisiene2007} studied by means of a quantitive spectral analysis are located in the ISM gas-rich parts of IC 1613. The same is true for the very few HII-regions, for which a direct determination of the oxygen abundance through a detection of auroral lines has been possible \citep{Bresolin2007}. This work will be challenging, though. There are many OB star associations in this region \citep{Garcia2009, Borissova2004}, but the members of these associations are relatively faint and long exposure times for optical or UV spectroscopy will be required. There are also many HII-regions \citep{Valdez2001}, but as pointed out by \cite{Bresolin2007} the general problem of quantitative HII-region spectroscopy in IC 1613 is that the surface brightnesses are low and the detection of auroral lines is difficult. This will require a dedicated study with very long exposure times, preferably with integral field spectroscopy at large telescopes.

An alternative and very promising spectroscopic method to obtain metallicities of massive young stars is medium resolution J-band technique of red supergiant stars developed by \cite{Davies2010} and \cite{Gazak2014}. This technique has already been applied to a variety of galaxies with very good results \citep{Gazak2015, Patrick2015, Patrick2017}. A differential study of red supergiants in the HI cavity and the surrounding ring area would be an excellent independent check of our blue supergiant result.

We note that in our discussion of star formation history and chemical evolution we have only considered the spatial distribution of neutral hydrogen in the ISM. If the central cavity would contain a significant amount of molecular hydrogen of much higher density than in in the surrounding ring, then our interpretation would have to be revised. At this point, we are not aware of deep enough molecular gas observations with sufficient spatial resolution, which would allow us to disprove or to confirm our proposed scenario.

The determination of a new independent distance to IC 1613 by means of the blue supergiant FGLR technique based on the new calibration by \cite{Urbaneja2017} yields good agreement with the Cepheid and the TRGB methods. It confirms the result by \cite{Hosek2014} that the FGLR method also seems to work at low metallicity. This is an encouraging result and further establishes the FGLR method as a new complementary tool to investigate the extragalactic distance scale. 

\acknowledgments
We thank the reviewer for the helpful and constructive comments. We also thank Ines Camacho, Miriam Garcia and Artemio Herrero for providing us with their results prior to publication and for a detailed discussion of our work. WG and GP gratefully acknowledge financial support for this work from the BASAL Centro de Astrofisica y Tecnologias Afines (CATA) PFB-06/2007. WG also gratefully acknowledges financial support from the Millennium Institute of Astrophysics (MAS) of the Iniciativa Cientifica Milenio del Ministerio de Economia, Fomento y Turismo de Chile, project IC120009.

\begin{deluxetable*}{cccccccccc}[h!]
\tabletypesize{\small}
\tablewidth{0pt}
\tablenum{1}
\tablecolumns{10}
\tablecaption{IC 1613 A Supergiants \label{tab:targ}}
\tablehead{
  \colhead{Star} & \colhead{\teff} & \colhead{\logg} & \colhead{\loggf} & \colhead{[Z]} &  \colhead{V} & \colhead{B - V} & \colhead{E(B-V)} & \colhead{BC} & \colhead{\mbol} \\
  \colhead{ } & \colhead{[K]} & \colhead{cgs} & \colhead{cgs} & \colhead{dex} & \colhead{mag} &  \colhead{mag} & \colhead{mag} & \colhead{mag} & \colhead{mag} \\
  \colhead{(1)} & \colhead{(2)} & \colhead{(3)} & \colhead{(4)} & \colhead{(5)} & \colhead{(6)} & \colhead{(7)} & \colhead{(8)} & \colhead{(9)} & \colhead{(10)}
}
\startdata
A1  & 8825 $\pm$ 175 & 1.95 $\pm$ 0.05 & 2.17$^{+0.10}_{-0.10}$ & -0.49 $\pm$ 0.11 & 18.800 &  0.061 & 0.084 $\pm$ 0.021 & -0.069 $\pm$ 0.038 & 18.429 $\pm$ 0.081 \\
A2  & 8125 $\pm$  70 & 1.62 $\pm$ 0.10 & 1.98$^{+0.11}_{-0.11}$ & -0.42 $\pm$ 0.07 & 19.443 &  0.204 & 0.198 $\pm$ 0.020 &  0.054 $\pm$ 0.010 & 18.843 $\pm$ 0.075 \\
A3  & 8300 $\pm$  70 & 1.40 $\pm$ 0.10 & 1.72$^{+0.11}_{-0.11}$ & -0.40 $\pm$ 0.07 & 18.543 &  0.287 & 0.275 $\pm$ 0.020 &  0.020 $\pm$ 0.010 & 17.655 $\pm$ 0.074 \\
A4  & 9000 $\pm$ 250 & 1.90 $\pm$ 0.05 & 2.08$^{+0.10}_{-0.12}$ & -0.70 $\pm$ 0.15 & 19.022 &  0.038 & 0.066 $\pm$ 0.021 & -0.129 $\pm$ 0.047 & 18.675 $\pm$ 0.088 \\
A4b &14700 $\pm$ 500 & 2.67 $\pm$ 0.10 & 2.00$^{+0.10}_{-0.10}$ & -0.45 $\pm$ 0.20 & 18.956 & -0.075 & 0.060 $\pm$ 0.021 & -1.141 $\pm$ 0.085 & 17.616 $\pm$ 0.114 \\
A6  & 8120 $\pm$  70 & 1.72 $\pm$ 0.07 & 2.08$^{+0.08}_{-0.09}$ & -0.60 $\pm$ 0.07 & 18.671 &  0.105 & 0.103 $\pm$ 0.020 &  0.043 $\pm$ 0.010 & 18.373 $\pm$ 0.075 \\
A8  & 9450 $\pm$ 150 & 1.14 $\pm$ 0.05 & 1.24$^{+0.05}_{-0.05}$ & -0.96 $\pm$ 0.07 & 16.445 &  0.122 & 0.090 $\pm$ 0.020 & -0.275 $\pm$ 0.043 & 15.873 $\pm$ 0.084 \\
A11 &10250 $\pm$ 200 & 1.64 $\pm$ 0.05 & 1.61$^{+0.05}_{-0.05}$ & -0.95 $\pm$ 0.07 & 18.062 &  0.008 & 0.034 $\pm$ 0.020 & -0.372 $\pm$ 0.039 & 17.577 $\pm$ 0.083 \\
B1  & 8400 $\pm$ 200 & 1.87 $\pm$ 0.10 & 2.17$^{+0.14}_{-0.15}$ & -0.93 $\pm$ 0.15 & 19.624 &  0.018 & 0.031 $\pm$ 0.022 & -0.028 $\pm$ 0.038 & 19.492 $\pm$ 0.086 \\
B8  &11700 $\pm$ 350 & 2.49 $\pm$ 0.05 & 2.22$^{+0.06}_{-0.06}$ & -0.50 $\pm$ 0.10 & 19.504 & -0.033 & 0.057 $\pm$ 0.021 & -0.617 $\pm$ 0.064 & 18.700 $\pm$ 0.098 \\
B9  & 8100 $\pm$ 200 & 1.90 $\pm$ 0.15 & 2.27$^{+0.20}_{-0.23}$ & -0.50 $\pm$ 0.15 & 19.707 &  0.137 & 0.131 $\pm$ 0.024 &  0.053 $\pm$ 0.038 & 19.328 $\pm$ 0.093 \\
B10 &12750 $\pm$ 350 & 2.56 $\pm$ 0.10 & 2.14$^{+0.10}_{-0.10}$ & -0.55 $\pm$ 0.15 & 18.918 & -0.052 & 0.052 $\pm$ 0.021 & -0.809 $\pm$ 0.062 & 17.963 $\pm$ 0.097 \\
B15 & 8300 $\pm$  70 & 1.60 $\pm$ 0.10 & 1.92$^{+0.11}_{-0.11}$ & -0.89 $\pm$ 0.07 & 18.779 &  0.067 & 0.075 $\pm$ 0.020 & -0.008 $\pm$ 0.010 & 18.234 $\pm$ 0.074 \\
B18 & 8650 $\pm$ 250 & 1.86 $\pm$ 0.10 & 2.11$^{+0.17}_{-0.16}$ & -1.25 $\pm$ 0.15 & 19.747 &  0.039 & 0.060 $\pm$ 0.021 & -0.087 $\pm$ 0.047 & 19.461 $\pm$ 0.088 \\
C1  & 8300 $\pm$  50 & 1.60 $\pm$ 0.05 & 1.92$^{+0.06}_{-0.06}$ & -0.85 $\pm$ 0.07 & 18.177 &  0.034 & 0.042 $\pm$ 0.020 & -0.006 $\pm$ 0.010 & 18.033 $\pm$ 0.073 \\
C2  & 8300 $\pm$ 400 & 1.95 $\pm$ 0.15 & 2.27$^{+0.24}_{-0.32}$ & -0.90 $\pm$ 0.30 & 19.630 &  0.104 & 0.117 $\pm$ 0.028 & -0.007 $\pm$ 0.076 & 19.237 $\pm$ 0.122 \\
C3  & 8000 $\pm$ 100 & 1.71 $\pm$ 0.15 & 2.10$^{+0.17}_{-0.17}$ & -0.85 $\pm$ 0.15 & 19.807 &  0.160 & 0.153 $\pm$ 0.020 &  0.051 $\pm$ 0.019 & 19.354 $\pm$ 0.075 \\
C8  & 8300 $\pm$ 350 & 1.70 $\pm$ 0.15 & 2.02$^{+0.24}_{-0.28}$ & -0.85 $\pm$ 0.35 & 19.891 &  0.122 & 0.133 $\pm$ 0.023 & -0.005 $\pm$ 0.067 & 19.477 $\pm$ 0.106 \\
C16 & 8200 $\pm$ 150 & 1.42 $\pm$ 0.10 & 1.76$^{+0.13}_{-0.14}$ & -0.45 $\pm$ 0.15 & 18.544 &  0.243 & 0.227 $\pm$ 0.021 &  0.043 $\pm$ 0.030 & 17.834 $\pm$ 0.080 \\
C17 & 8600 $\pm$ 200 & 2.12 $\pm$ 0.10 & 2.38$^{+0.14}_{-0.14}$ & -0.55 $\pm$ 0.15 & 19.270 &  0.074 & 0.094 $\pm$ 0.022 & -0.044 $\pm$ 0.038 & 18.914 $\pm$ 0.088 \\
C18 &12900 $\pm$ 500 & 2.68 $\pm$ 0.10 & 2.24$^{+0.10}_{-0.10}$ & -0.45 $\pm$ 0.25 & 19.577 & -0.053 & 0.057 $\pm$ 0.021 & -0.829 $\pm$ 0.088 & 18.559 $\pm$ 0.117 \\
\enddata
\tablecomments{Object identifications from \cite{Bresolin2007}.}
\end{deluxetable*}

\bibliography{IC1613-BSGs}

\end{document}